\documentclass[%
reprint,
amsmath,amssymb,
raggedbottom,
aps,
floatfix,
showkeys,
prm, 
superscriptaddress]{revtex4-2}

\usepackage{booktabs}
\usepackage{multirow}
\usepackage{graphicx}
\usepackage{bm}
\usepackage{mathtools}
\usepackage{cleveref}
\usepackage{amsmath}
\usepackage{ulem}

\usepackage[utf8]{inputenc}
\usepackage[T1]{fontenc}
\usepackage{mathptmx}
\usepackage{etoolbox}
\usepackage{physics}
\usepackage{color}
\usepackage[font=small,skip=3pt]{caption}
    \usepackage{hyperref}
    \hypersetup{colorlinks=true,
                allcolors=black,   
                filecolor=blue,
                citecolor=blue,
                linkcolor=blue,
                urlcolor=blue,
                menucolor=black,
                anchorcolor=black,
                }
	\usepackage{float}
	\usepackage{lipsum}
\begin{document}

\preprint{Number of Preprint}

\title[Title]{ Deformation assisted precipitation in binary alloys:\\ A competition of time-scales}
\author{Alex Mamaev}
	\email{Author to whom correspondence should be addressed. E-mail: alexandre.mamaev@mail.mcgill.ca}
    \affiliation{Department of Physics, Centre for the Physics of Materials, McGill University, Montreal, QC, Canada H3A 2T8}
    \author{Duncan Burns}
    \affiliation{McCormick School of Engineering, Northwestern University, Evanston, IL, United States 60208}
	\author{Nikolas Provatas}
	\affiliation{Department of Physics, Centre for the Physics of Materials, McGill University, Montreal, QC, Canada H3A 2T8}

\date{\today}

\begin{abstract}
We consider the process of precipitation in binary alloys in the presence of mechanical deformation. It is commonly observed that mechanical deformation prior to or during precipitation leads to microstructure with excess defects, which allows for enhanced precipitate nucleation and growth rates \cite{Weiss1979,Okaguchi1992,Deschamps2003}. To investigate this phenomenon, we employ a two-dimensional phase-field crystal alloy model endowed with a temperature dependent mobility, making it capable of recovering isothermal transformation (TTT) diagrams with a characteristic inflection point (nose) about a critical temperature. We examine the variation in the time-scale of precipitation and its connection to the time-scale of the applied deformation, focusing on the roles of atomic defects in the processes involved. Our results indicate that precipitation is initially delayed through application of a deformation until a critical strain is achieved, beyond which  precipitation proceeds more rapidly, assisted by plastic deformation such as grain boundary serration or dislocation nucleation. We show that the evolution of the precipitated fraction, $f(t)$, departs from classical Avrami behaviour. Specifically, $df/dt$ develops two peaks indicative of a ``plateau"-like inflection in $f(t)$, signalling the transition to defect assisted precipitate nucleation. We analyze these plateaus as a function of the deformation rate and demonstrate that the they exhibit a discontinuous bifurcation as the time-scale of applied deformation is increased. These findings are compared to and found to be consistent with experiments.
    
\end{abstract}
\maketitle
\section{Introduction}
Mechanical deformations driving out-of-equilibrium behaviours are ubiquitous in current and emerging material processing techniques such as additive manufacturing, laser shock peening, friction stir-welding, etc. \cite{Dhakal2020, Pan2021_2, Liao2010}.

The material conditions left behind in the wake of these processes provide inflated driving forces for diffusive and phononic relaxation due to increased saturation of solid solutions and the presence of point defects. This is in sharp contrast to classical casting conditions \cite{Zhakhovsky2023, Budzevich2012}. An emphasis on the physics governing the aforementioned material processes has thus been of interest to the materials science community, particularly how such processes deviate from classically understood mechanisms of precipitation, recrystallization and recovery in metal alloys.

Precipitation readily occurs at sites of higher energy, such as defects and Grain Boundaries (GBs) \cite{Cahn1957}. Mechanical deformation of crystalline solids produces many such sites, thus modifying the nucleation, growth, and coarsening properties encountered \cite{Deschamps2003}.

Precipitation can thus be seen as competitive with other forms of strain relaxation eg. recrystallization and recovery in metals and their alloys. These processes all use stored driving forces to relieve stress accruing at high energy sites within a deformed sample. Indeed, thus has been studied previously \cite{Weiss1979, Okaguchi1992}, where experiments and numerical simulations have examined modifications to isothermal transformation (TTT) diagrams in the presence of mechanical deformation. Analytical models have likewise been formulated to quantify this interactive growth process \cite{Zurob2001}. These models typically require a priori knowledge of specific material properties which are difficult to extract such as activation energies and defect fractions. Furthermore, these models require knowledge of the nucleation and growth mechanisms of both recrystallized and precipitated grains.  

Recent indirect experimental measurements, such as in-situ small-angle X-ray scattering of deforming Al-Zn-Mg-Cu, have provided detailed views of the kinetics of precipitation \cite{Deschamps2003, Ma2021}. These studies have demonstrated an enhanced precipitation, in particular, an enhanced precipitated volume fraction growth rate is observed upon application of deformation, argued to be a result of an increased defect density. Despite these advances, direct in-situ observation of microstructure evolution is typically impossible. Hence, the nano-scale mechanisms involved in the formation of precipitates cannot be thoroughly examined experimentally, and many questions about defect-assisted precipitation remain. Due to the above challenges, the development and application of numerical modelling methods to examine the nano-scale behaviour during such processes becomes crucial.

This paper is structured as follows: In Section \ref{sec:PFC}, we introduce the details of the XPFC model we employ including temperature dependent mobilities. In Section \ref{sec:Numerics}, we describe the numerical approach we take to model a precipitating system under continuous deformation. In section \ref{sec:Results}, we present our results, including the demonstration of a second, deformation induced time-scale in the dynamics of precipitation. Finally in Section \ref{sec:Conclusion}, we briefly provide a future outlook and describe how our results may be interpreted.

\section{Phase-Field Crystal Model}\label{sec:PFC}

The phase-field crystal (PFC) technique is an atomistic modelling approach for microstructure evolution that, unlike traditional phase-field modelling, is capable of simulating the coupled evolution of solute diffusion and the effects of lattice orientation, grain boundaries, dislocations, and thermal and solute based strain effects on experimentally relevant length and time-scales \cite{Elder2004, Elder2007, Ofori2013, ProvatasBook, Greenwood2010,Berry2006,Berry2014_2,Burns2024_1}. Moreover, by introducing a second time-scale into the density evolution, PFC models may also support quasi-phononic relaxation while remaining orders of magnitude faster than molecular dynamics simulations \cite{Stefanovic2006,Stefanovic2009,Burns2022}. This endows PFC models with the ability to effectively examine the interplay between recrystallization and precipitation.

\subsection{Free Energy Functional}\label{sec:PFC_Energy}
In order to investigate the atomistic behaviour of precipitation in the presence of deformation, this work examines the prototypical scenario of an equal tilt bicrystal undergoing a continuous biaxial deformation with strain rate parameter $\dot{\epsilon}$, which shall be clarified later. To this end, we employ a structural phase field crystal (XPFC) model with a free energy functional that models substitutional binary alloys and has previously been applied to studies involving precipitation \cite{Fallah2012,Fallah2013}, as well as deformation in such systems \cite{Chensuang2023}. Our free energy is formulated as follows, 
\begin{align}
\label{eq:free_energy}
    \begin{split}
    \frac{\Delta F}{k_B T^0\rho^0}= 
    \tau&\int{d\mathbf{r}}\bigg\{
    \frac{1}{2}n^2-\frac{\eta}{6}n^3+\frac{\chi}{12}n^4
    +\omega \,\Delta F_{mix} \\
    +&\frac{W_c}{2}|\nabla c|^2 -\frac{1}{2}n\int{d\mathbf{r}^\prime}\,
    C_{nn}^{(2)}(|\mathbf{r}-\mathbf{r}^\prime|) \, 
    n^\prime \bigg\},
    \end{split}
\end{align}
where $n=(\rho-\rho^0)/\rho^0$ measures the local coarse grained density relative to a reference density, $\rho^0$ corresponding to the expansion point of the model, $\tau=T/T^0$ is a dimensionless temperature and $c$ is the local solute concentration field. Here, $\eta$ and $\chi$ control deviations from the ideal gas free energy, $W_c$ controls the diffusion length of the $c$ field and $\omega$ controls the free energy of mixing, given by
\begin{align}
    \Delta F_{mix}= &(1+n)\bigg\{c\log(\frac{c}{c_0})+(1-c)\log(\frac{1-c}{1-c_0})\bigg\},
    \label{mixing_energy}
\end{align}
where $c_0$ is the concentration at the reference. The final,  excess, term of the free energy controls the length scale and strength of atomic-scale interactions. The interactions are embedded in the two-point density-density correlation function $C_{nn}$ which interpolates between pure $A$ and pure $B$ structures through compositionally dependent functions $\chi_A(c) = 1-3c^2+2c^3$, $\chi_B(c) = 1-3(1-c)^2+2(1-c)^3$ such that
$ C_{nn}^{(2)}(\mathbf{r}-\mathbf{r'}) = \chi_A C_{AA}^{(2)}(\mathbf{r}-\mathbf{r'}) + \chi_B C_{BB}^{(2)}(\mathbf{r}-\mathbf{r'})$. In XPFC, the pure species correlation functions, $C_{ii}^{(2)}$ are defined in Fourier space (with wave-vector magnitude $k$) by Gaussian functions centered about the dominant reciprocal lattice wave-vector ($k=k_{i}$) and $k=0$ modes of each species respectively,
\begin{align}\label{eq:correlation}
    \hat{C}_{ii}^{(2)}(k) =e^{-\frac{T}{T_i^0}} \bigg\{-A_0e^{-\frac{k^2}{2\sigma_{i,0}^2}}+e^{-\frac{(k-k_i)^2}{2\sigma_{i,1}^2}}\bigg\},
\end{align}
where $i=A,B$. The parameter $T_i^0$ controls the temperature scaling of the corresponding correlation function,  $\sigma_{i,0}$ and $\sigma_{i,1}$ control the Gaussian widths of the $k=0$ and dominant wave-vector respectively. Finally, $A_{i,0}$ is a constant that controls the amplitude of the $k=0$ mode. In this letter, we work in two spatial dimensions and employ a single wave-vector in our correlation to stabilize a hexagonal structure in each solid phase. The details of the parameters we employ and the resulting phase-diagram are shown in Appendix {\color{blue} \ref{sec:Appendix_A}}.
\subsection{Dynamics}\label{sec:PFC_Dynamics}
To evolve the $n$ and $c$ fields, we apply conserved dynamics,
\begin{align}
     \frac{\partial^2 n}{\partial t^2}+ \beta \frac{\partial n}{\partial t} & = \nabla M_n \cdot \nabla \mu_n +\xi_n,\label{eq:n_dynamics} \\
    \frac{\partial c}{\partial t} & = \nabla M_c \cdot \nabla \mu_c+\xi_c,\label{eq:c_dynamics}
\end{align}
where $\mu_{n/c}$ are the chemical potentials, $M_{n/c}$ are the mobilities and $\xi_{n/c}$ are noise sources applied to the $n$ and $c$ fields respectively. The noise terms satisfy the fluctuation dissipation relations, $\langle\xi_{n/c}(\textbf{r},t)\xi_{n/c}(\textbf{r'},t')\rangle= N_{n/c}M_{n/c}k_BT\delta(\textbf{r}-\textbf{r'})\delta(t-t')$ where $N_{n/c}$ are model parameters subsuming remaining constant prefactors. The noise terms here are implemented to allow for stochastic variations between different realizations of the same initial conditions. Without such fluctuations, the driving force for precipitation presented by enhanced solute segregation to dislocations and GBs may not be sufficient to nucleate precipitates despite a quench into the metastable region. In such systems, fluctuations act to drive the concentration field over the barrier present.

We also add an additional time-scale through the parameter $\beta$ for quasi-phonon relaxation into the density equation. Here, $\beta$ defines the effective length-scale over which quasiphonons propagate undamped\cite{Stefanovic2006}. A two-time scale coupling to $c$ is not examined here as we do not anticipate any transport of solute on time-scales commensurate with elastic relaxation.

It is worthwhile to note that Eqs. (\ref{eq:n_dynamics}) and (\ref{eq:c_dynamics}) are not strictly consistent in the regime where the system's average density has significant variations \cite{Jugdutt2015}. A more rigorous treatment of the dynamics has been implemented in \cite{Frick2020}. However, as the common tangents between the free energy wells of our system exist at small and nearly identical average densities, this approximation is justified. 

We expand on the model by introducing temperature dependent mobilities into Eqs. (\ref{eq:n_dynamics}) and (\ref{eq:c_dynamics}). This allows for the generation of TTT diagrams that can be controlled through model parameters. An Arrhenius form is assumed for the mobilities of both $n$ and $c$,
\begin{equation}
    M_{n/c} = M_{n/c}^0 e^{-E_A^{n/c}/k_BT}\label{eq:mob_nc}
\end{equation}
where $E_A^{n/c}$ and  $M_{n/c}^0$ are the activation energies, and the mobilities at $T=0$ respectively, which control the curvature and position of the inflection point of a given isothermal transformation (TTT) curve for precipitation.
Formally, the mobilities $M_{n/c}$ are non-linearly coupled to the mobilities of the individual atomic species. In this paper, we treat them as mutually independent model parameters. It is also noted that the modified mobilities will shift the 
quasiphonon propagation decay length. No qualitative differences are observed in our results due to such shifts.

\begin{figure}[t]
    \centering
    \includegraphics[width = \columnwidth]{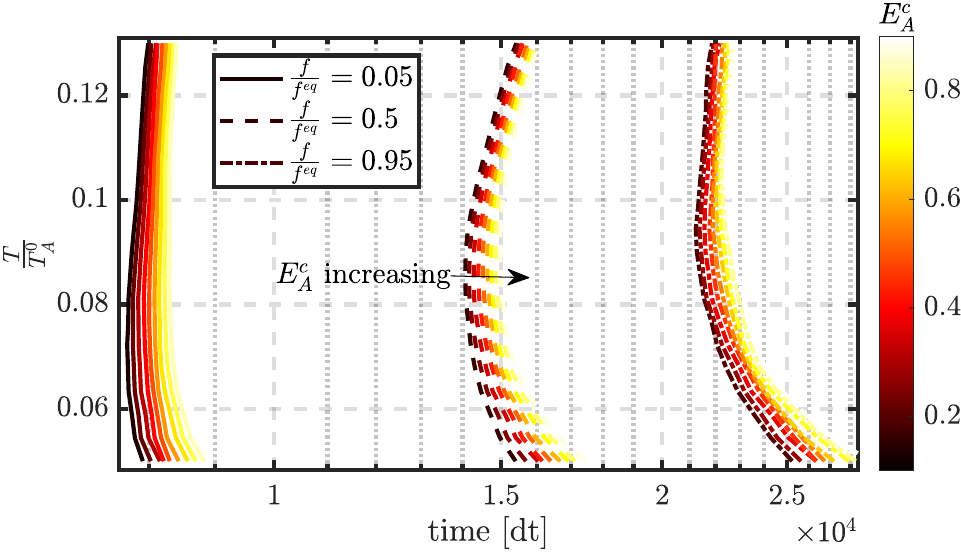}
    \caption{The TTT diagram of precipitation at zero strain rate. The three sets of curves represent different fractions of completed transformation. A characteristic inflection or nose shape can be observed over the range of temperatures selected as a result of the temperature dependent mobilities defined in Eq.~(\ref{eq:mob_nc}). Curves are plotted for a selection of $E_A^c$, which monotonically controls the reaction rate as shown. }
     \vspace{-1.8em}
    \label{fig:TTT}
    
\end{figure}

\begin{figure*}
    \centering
    \includegraphics[width = \textwidth]{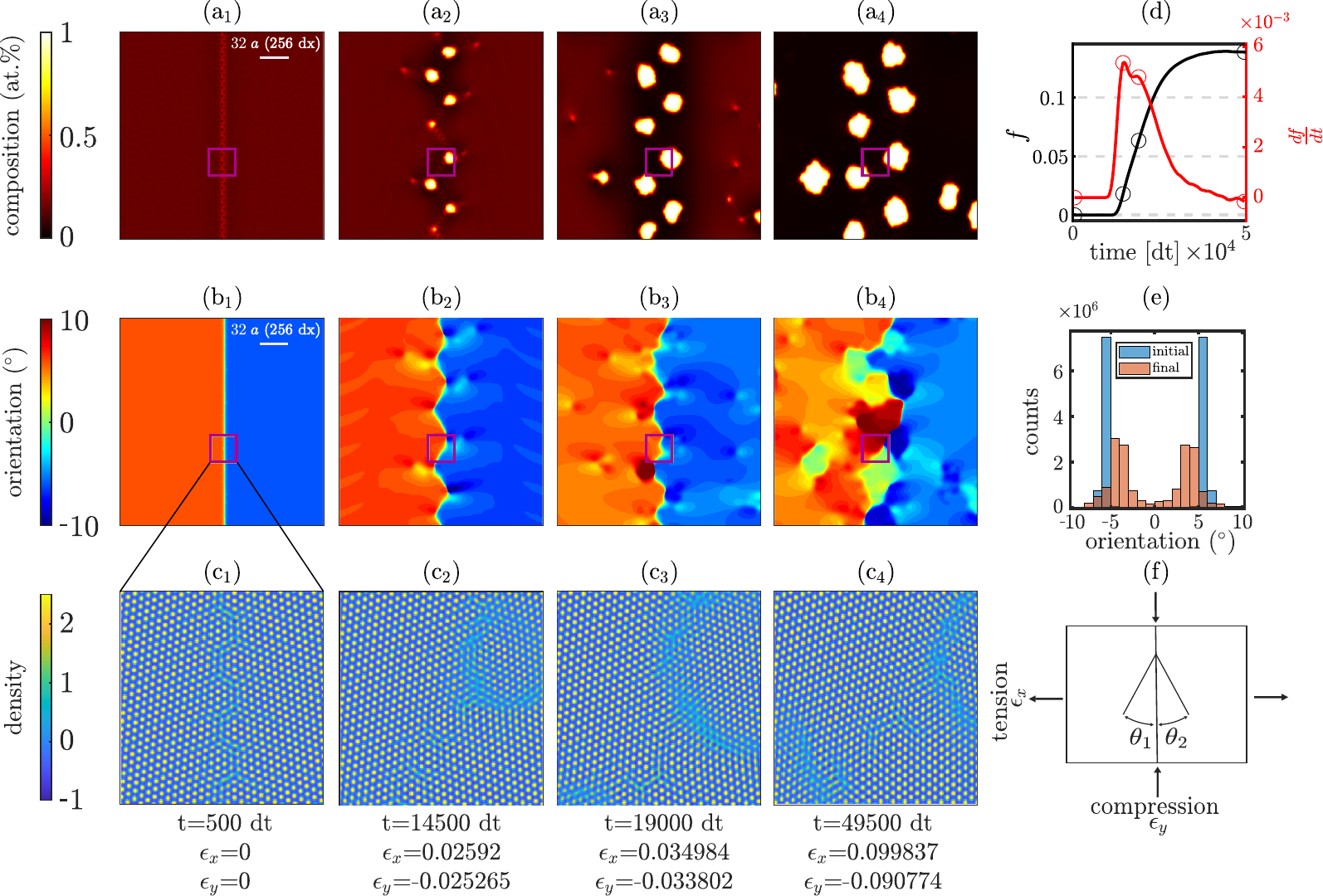}
    \centering
    \vspace{-0.8em}
    \caption{The sequence of events of a simulation of strain enhanced precipitation. Panels ($\text{a}_{1}$)-($\text{a}_4$) show the evolution of the composition field and panels ($\text{b}_{1}$)-($\text{b}_4$) show the behaviour of the local orientation field. Panels ($\text{c}_{1}$)-($\text{c}_4$) show a zoom-in of the density field in the highlighted magenta regions of panel rows ($\text{a}$) and ($\text{b}$). The times and strains along each axis associated with each column of panel rows (a), (b) and (c) are shown at the bottom of the figure. Panel (d) shows the associated evolution of the precipitated fraction $f$ as well as $df/dt$ in black and red, respectively, with the values corresponding to panels ($\text{a}_{1}$)-($\text{a}_4$) labelled with circles. Panel ($\text{e}$) shows the histogram of orientations of panels ($\text{b}_1$) and ($\text{b}_4$). Panel ($\text{f}$) shows a schematic of our system including the directions of tension and compression. Our tilt axis is parallel to the y axis and we consider an axisymetric tilt where $\theta_1\approx -6^{\circ}$ and $\theta_2 \approx +6^{\circ}$. The first peak in ${df}/{dt}$, corresponding to panels ($\text{a}_2$), ($\text{b}_2$) is associated with the precipitation on the serrated interface. The second peak, corresponding to panels ($\text{a}_3$), ($\text{b}_3$) coincides with precipitate nucleation on dislocations. Panels ($\text{b}_4$) and ($\text{c}_4$) show the late stage coarsening behaviour of the system, which includes the reorientation of A-rich regions close to the grain boundary down to orientations intermediate between zero and the initial orientation of $\pm 6^{\circ}$.
     }
     \vspace{-1.5em}
    \label{fig:dynamics}
\end{figure*}

\section{Numerical Simulation Details\label{sec:Numerics}}
For all simulations described, we performed a quench below the solvus line into solid-solid coexistence. Full simulation parameters are presented in the Supplementary Material. We initialize the density field of the crystalline regions by seeding in an equal tilt bicrystal through a one-mode expansion. The lattice constant of this solid solution is set to be commensurate to the value attained by the interpolated correlation function for the given choice of supersaturation. The system is set to be initially strain-free by performing a minimization over the angle and number of atoms in the x and y directions for one (and by symmetry the other) grain as in \cite{Mellenthin2008}. The initial misorientation during this minimization was kept within the range 12.00$\pm 0.05^\circ$. The average density and supersaturation are set to be initially uniform and equal to 0.05 and 0.15, respectively. 

The density field is next allowed to relax by setting $M_c$ to zero. This relaxation entails the evolution of only Eq. (\ref{eq:n_dynamics}) until the amplitude of the density field reaches an equilibrated value in the bulks of the grains. After this time, Eq. (\ref{eq:c_dynamics}) is solved concomitantly with Eq. (\ref{eq:n_dynamics}) introducing solute dynamics. At this stage, an isochoric biaxial deformation is applied to the system. This is done by varying the system discretization along both axes, similarly to the approach employed in \cite{Chensuang2023}. We note that the strain rate parameter $\dot{\epsilon}$ is higher than the true strain rate by approximately two orders of magnitude due to the numerical methodology employed for deformation. In the proceeding discussion, for convenience, we will nevertheless refer to $\dot{\epsilon}$ as the strain rate. We provide further numerical details in Appendix \ref{sec:Appendix_B}.

Over the course of simulations we track the grand potential energy (analogously, the dimensionless system pressure $\bar{P}=P/k_BT^0\rho^0V$ \cite{Kocher2015}), precipitated fraction $f$, and local orientation. The orientation is calculated as in \cite{Wang2013} and $f${\color{red}\sout{,}} is calculated through thresholding in composition in conjunction with a Fourier space mask sharply peaked around the dominant wave-vector of the precipitating phase.

\section{Results \& Discussion\label{sec:Results}}
We first examine the situation where $\dot{\epsilon}=0$. While precipitation has been studied with XPFC in the past, to our knowledge no study has been performed with temperature dependent mobilities as in Eq.~(\ref{eq:mob_nc}). Over the course of each isothermal simulation, fluctuations drive the formation of precipitates along the GB of the system.
Over all temperatures and activation energies examined, the precipitate fraction obeys an Avrami-like behaviour 
\begin{equation}
    f = f^{eq}\bigg(1-e^{-\big(\frac{t}{\tau_{0.5}}\big)^m}\bigg),
\end{equation}
where $m$ is an exponent determined by the dimensionality and mechanism of the transformation, $f^{eq}$ is the equilibrium precipitate fraction and $\tau_{0.5}$ is a characteristic time-scale. The system pressure was found to exhibit a single dominant peak corresponding to the end of precipitate nucleation. Beyond this time, the system reduces its free energy and pressure through diffusion of solute out of the surrounding supersaturated bulk and into precipitate growth. 

Figure~\ref{fig:TTT} shows the TTT curves for a range of model temperatures, demonstrating the inflection about a critical temperature. The scaling of this surface is controlled by the usual driving force for precipitation (competition between bulk and surface energies) as well as the new mobility used here, which also affects the noise strength. At each temperature, the bulk energies are determined through the curvatures of the free energy landscape about the equilibrium wells of Eq.~(\ref{eq:free_energy}) in the space of average density, supersaturation and amplitude of the atomic density field $n$. The surface energy will be dependent on all of the above along with the effective gradient coefficient of the amplitude arising from the excess energy term, as well as $W_c$. These parameters are all held fixed for each value of $T$ along the curves shown in Figure~\ref{fig:TTT}. 

We next examine the behaviour of precipitation in the presence of mechanical deformation. We consider a single temperature below with the same initial conditions. 

The evolution of a precipitating system under mechanical deformation is shown in Figure \ref{fig:dynamics}. At lower values of $\dot{\epsilon}$, $f$ obeys the same qualitative Avrami behavior as the undeformed case. Prior to precipitation, in order to compensate for the applied deformation, the density field undergoes elastic accommodation. The characteristic time-scale, $\tau_{0.5}$, for the precipitation is observed to initially increase with $\dot{\epsilon}$. This occurs due to an accelerated rotation of the bicrystal with increasing $\dot{\epsilon}$, causing a reduction in the effective driving force for precipitation along the boundary. The increase we observe in $\tau_{0.5}$ follows a simple $\exp(-\Delta\gamma^2)$ dependence in the time-scale for precipitation. Here $\Delta\gamma = \gamma_f-\gamma_i$ where $\gamma_f$ is the precipitate-matrix surface energy and $\gamma_i$ is the initial surface energy, whose misorientation dependence has been shown to reproduce the correct Read-Shockley scaling \cite{Stolle2014}. We have also performed simulations in the undeformed case with varying orientation, recovering the same Read-Shockley scaling.

\begin{figure}[t]
    \centering
    \includegraphics[width = 0.9\columnwidth]{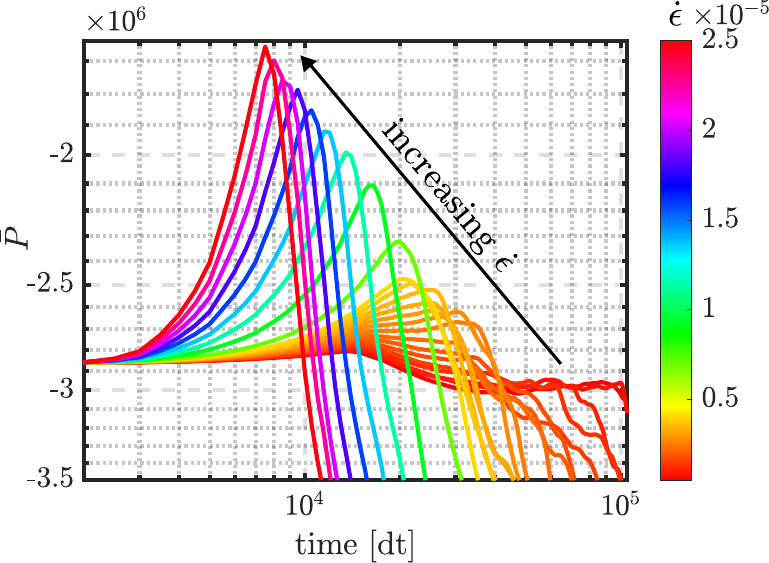}
    \caption{The system pressure is plotted as a function of system time for different choices of the strain rate. At lower values of $\dot{\epsilon}$ (dark red), an initial peak followed by oscillations indicating nucleation of dislocations post-precipitation is observed. At larger $\dot{\epsilon}$ (orange-yellow) a second peak emerges corresponding to a new dislocation assisted precipitation mechanism. At the largest range of $\dot{\epsilon}$ simulated (blue-pink), a single peak emerges again indicating precipitation assisted through GB serration. }
    \label{fig:Pressure}
\end{figure}

Figure~\ref{fig:Pressure} plots the system pressure as a function of time at different strain rates $\dot{\epsilon}$. As in the zero deformation case, at low $\dot{\epsilon}$ the pressure is first found to exhibit a single dominant peak at the initial stages of precipitation, followed by oscillations indicating accommodations by the GB and the formation of any dislocations necessary to support the growth of the population of precipitates.
As $\dot{\epsilon}$ is increased further beyond $2 \times 10^{-6}$, we observe three mechanisms that contribute to a systematic deviation in the precipitation mechanism, each activating in order of increasing $\dot{\epsilon}$.
\begin{figure}[t]
    \centering
    \includegraphics[width = 0.9\columnwidth]{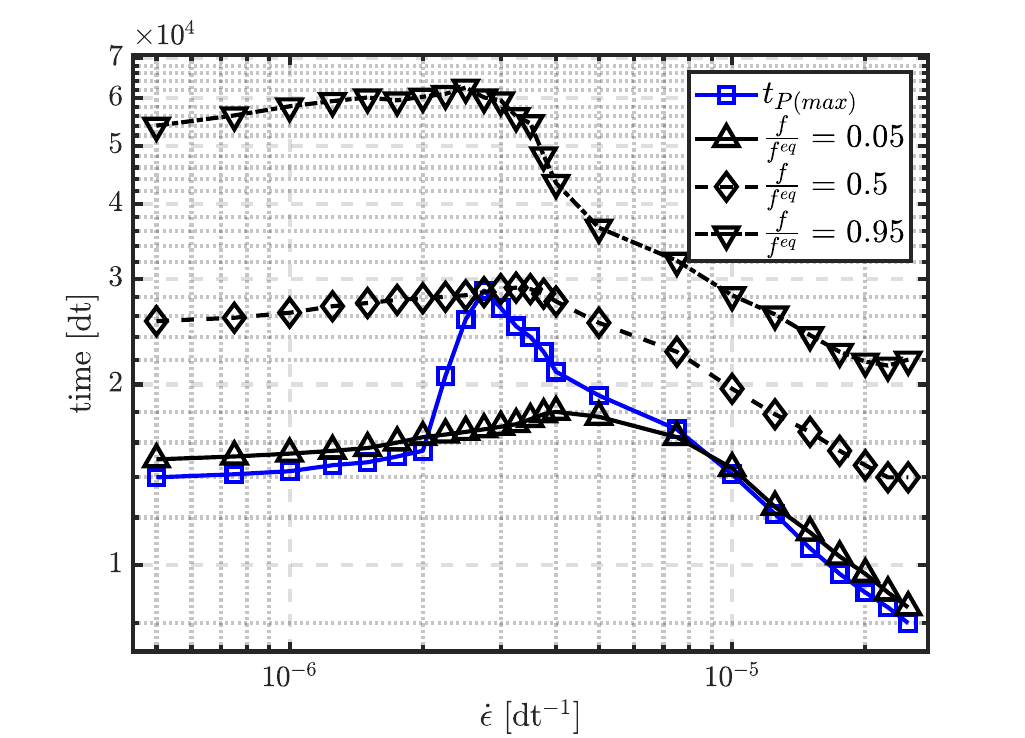}
    \caption{The times when $f/f^{eq}$ attains $0.05,0.50$ and $0.95$ plotted against strain-rate. Also shown in blue is the time at which the system achieves its maximum in pressure $t_{P(max)}$ at each strain-rate. At low $\dot{\epsilon}$, $t_{P(max)}$ corresponds to the initialization of precipitation. At a critical $\dot{\epsilon}=2\times 10^{-6}$,  $t_{P(max)}$ diverges significantly from its value during the initiation of precipitation. This occurs because the prominence of the second, plasticity induced, pressure peak surpasses the peak associated with precipitation, as shown in Figure \ref{fig:Pressure}. Upon further increase of $\dot{\epsilon}$ beyond $5\times10^{-6}$, serration of the GB occurs and $t_{P(max)}$ monotonically decreases past its initial value.}
    \label{fig:Strained_Time_Scales}
\end{figure}

\begin{figure*}
    \centering
    \includegraphics[width = 0.9\textwidth]{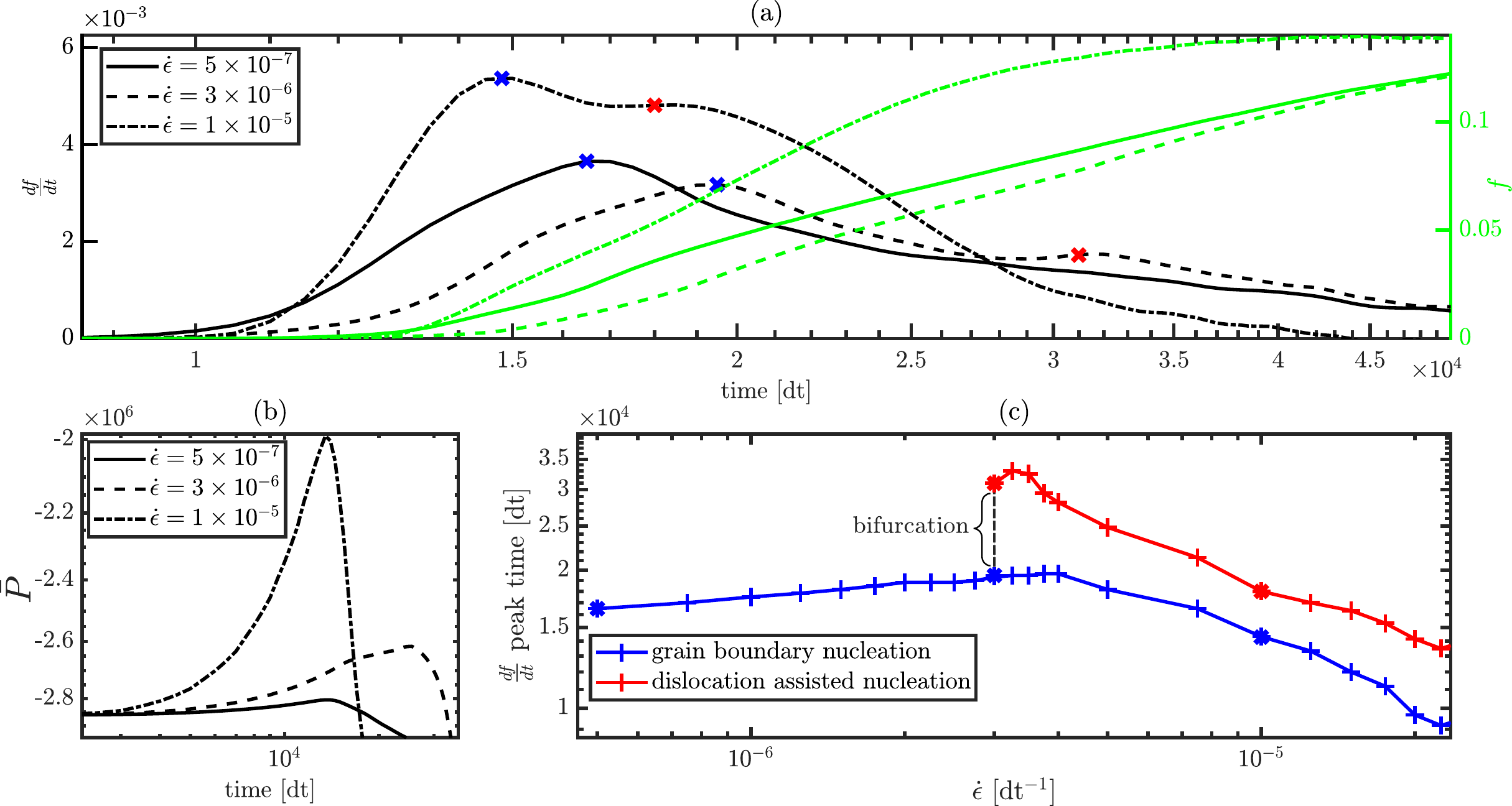}
    \caption{Panel (a) shows $f$ (green) and $df/dt$ (black), indicating the first and second peaks of $df/dt$ (blue and red respectively). The data are plotted for three values of progressively larger $\dot{\epsilon}$. The low $\dot{\epsilon}$ scenario demonstrates a single peak whereas the second and third cases exhibit a prominent second peak in $df/dt$, associated with dislocation nucleation. Panel (b) shows the pressure changes associated with these three choices of $\dot{\epsilon}$. In panel (c), blue data show the first peak in $df/dt$ vs. $\dot{\epsilon}$. This corresponds to the nucleation of precipitates on the GB. Beyond a critical strain rate of $\dot{\epsilon}=3\times 10^{-6}$ dt$^{-1}$, the second peak in $df/dt$ appears, shown here in red. } 
    \vspace{-1.5em}
    \label{fig:Peak_Time}
\end{figure*}

At the lower range of $\dot{\epsilon}$, we observe nucleation of dislocations from initial precipitates or the GB, also shown in Figure \ref{fig:dynamics}. It is well known that dislocations act as preferential nucleation sites for precipitation \cite{Cahn1957}. This has also been examined numerically in prior studies using the XPFC model \cite{Fallah2012, Fallah2013}. Excess solute may diffuse to not only the GB but also towards the dislocations flowing into the bulk of the system, inducing further precipitation at sites away from the GB. Dislocation nucleation events are characterized by a new peak in the system pressure that progressively increases in prominence and occurs after initial precipitation along the GB. This peak is subsumed into the pressure shift due to precipitation at intermediate $\dot{\epsilon}$ as shown in Figures~\ref{fig:Pressure} and~\ref{fig:Strained_Time_Scales}. The same effect also leads to the emergence of a second peak in $df/dt$, shown in Figure~\ref{fig:Peak_Time}, corresponding to additional precipitate nucleation mechanism.
Upon further increase in $\dot{\epsilon}$, deformation causes the GB to serrate as shown in Figure \ref{fig:dynamics}. This serration provides preferential sites of precipitate nucleation due to shifted contact angle and curvature induced solute segregation. Furthermore, formation of these precipitates relieves the strain energy stored in the serrated GB. As shown in Figure~\ref{fig:Pressure}, the resulting pressure for high values of $\dot{\epsilon}$ exhibits a single prominent peak, accounting for both a buildup through this serration mechanism and subsequent release from precipitation.

A third mechanism observed at high $\dot{\epsilon}$ is recrystallization at the GB, ie. the formation of lower-strain A-rich grains. These grains further serve to lower the barrier for precipitate nucleation. Similar behaviour has recently been modelled in a pure material PFC model under laser deposition \cite{Duncan2024,Burns2023}. A typical view of recrystallization events we observe are shown in the orientation field of Figure \ref{fig:dynamics}. Due to working in two spatial dimensions these recrystallized regions are not long lived. Indeed, upon performing identical simulations in the scenario where solute is frozen ($M_c,\xi_c =0$), we see the recrystallized regions initially attain an orientation intermediate to the surrounding bulk grains and eventually coarsen, post nucleation. The mechanism of formation for these recrystallized grains is achieved through collective motion of nucleated dislocations to serrated regions of the GB. In the absence of precipitation, this mechanism has been observed experimentally in the nucleation of twins during continuous deformation of copper bicrystals \cite{Miura2004}.

To summarize, the collective action of the above three phenomena causes $f$ to exhibit a plateau-like behaviour. As shown in Figures \ref{fig:dynamics} and \ref{fig:Peak_Time}, the maximum in $df/dt$ discontinuously transitions from one to a pair of discrete maxima. Such behaviour in Avrami curves is commonly associated with multiple transformation mechanisms of similar driving forces competing with one another.

Our findings have been observed experimentally in the situation when recrystallization is halted by precipitation \cite{Medina1997} and in the precipitate fraction of an Al-Zn-Mg-Cu  during hot deformation \cite{Deschamps2003}. 
These experimental results have also been examined from the perspective of a Kampmann-Wagner model equipped with a vacancy concentration and an evolving dislocation density \cite{Bignon2022}. However, their results do not consider nucleation of precipitates on dislocations. Experimentally, such dislocations are expected to be released during simultaneous deformation and precipitation \cite{Embury2003}.
It is noted, however, that strain-rate dependence in the present study cannot be quantitatively compared to the mechanisms proposed in \cite{Deschamps2003} as defects cannot frustrate in two dimensions. Nonetheless, our results suggest that an additional strain-induced dislocation nucleation mechanism may be required to fully model the dynamics of precipitation.

Our findings also indicate that the continuous application of a deformation may induce an acceleration of the precipitation reaction. This is because the introduction of defects enhances precipitate nucleation in the bulk. As shown in the data of Figure \ref{fig:Strained_Time_Scales}, precipitation under large deformation occurs over a shorter time-scale than the case of low (or no) deformation rate, beyond which we observe a bifurcation in $df/dt$. These results are also consistent with typically observed experimental behaviour of TTT diagrams in the presence of deformation \cite{Weiss1979,Chen2020}.

\section{Conclusion\label{sec:Conclusion}}
To conclude, we have shown that modified mobilities introduced into an improved XPFC alloy model as in Eq.~(\ref{eq:mob_nc}) provide a robust atomistic-continuum framework for examination of first order phase transformations in the solid state. 
This framework allows for temperature based control of the activity of structural vs. diffusively driven processes that may be tuned independently to driving forces arising from the free energy.

We have further shown that upon application of deformation to a bicrystal, beyond an initial stage of elastic compensation where precipitation occurs along a flat GB, new defect assisted nucleation mechanisms begin to compete with classic precipitation mechanisms along GBs. Our results demonstrate that these mechanisms alter the dispersion of precipitates within the system and introduce a second time-scale into the Avrami dynamics of these systems. 

The results of this study demonstrate how PFC modeling may be used to elucidate how precipitate matrix interactions on multiple time and length scales affect microstructure evolution. In regards to hot deformation, information about atomistic properties such as the dislocation density and GB behaviour can be further investigated. Additionally, future studies in three spatial dimensions may be able to shed further light on both recrystallization and defect pileups in such systems.

\acknowledgments
NP thanks The Natural Sciences and Engineering Research Council of Canada (NSERC) and the Canada Research Chairs (CRC) Program for funding support, and the Digital Research Alliance of Canada for computing resources. 
\appendix
\section{Free Energy Parametrization\label{sec:Appendix_A}}
\begin{figure}[]
    \centering
    \includegraphics[width = 0.9\columnwidth]{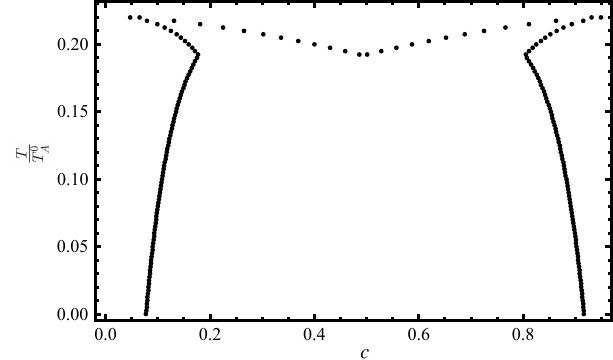}
    \caption{Phase diagram associated with free energy parameters shown in Table \ref{tab:PD_Params} with respect to dimensionless temperature, $T/T^0$ and concentration $c$. The average density is fixed here at 0.05 which is sufficiently distant from any phases outside of the liquid and solid phase.}
    \label{fig:phase_diagram}
\end{figure}

The parameters we employ that specify the ideal, mixing and gradient components of the free energy in Equation \ref{eq:free_energy} to our particular material of interest are shown in Table \ref{tab:PD_Params}. The parameters we employ for the correlation function in Equation \ref{eq:correlation} are shown in Table \ref{tab:Correlation_Params}. These parameters are held fixed over the course of every simulation we perform. The respective phase diagram corresponding to these parameters is also shown in Figure \ref{fig:phase_diagram}. Phase diagram construction is performed by numerically identifying common tangents in the free energy landscape after an assumed mode expansion for the density is inserted into the free energy and a subsequent minimization of the mode expansion amplitudes is performed as described in {\it c.f.} \cite{Greenwood2010,ProvatasBook}.
\begin{table}[h]
    \centering
    \begin{tabular}{|c|c|c|c|c|}
     \hline
         $c_0$&    $\eta$ &
    $\chi$ &$\omega$ & $W_c$ \\
    \hline
    0.5 & 1.4 & 1 & 0.005 & 0.25 \\
    \hline
    \end{tabular}
    \caption{Free energy parameters employed in Eq. (\ref{eq:free_energy})}
    \label{tab:PD_Params}
\end{table}
\begin{table}[h]
    \centering
    \begin{tabular}{|c|c|c|c|c|c|c|c|c|c|}
     \hline
          $k_A$ & $k_B$ & $T_A^0$ & $T_B^0$ & $\sigma_{A,0}$&$\sigma_{B,0}$&$\sigma_{A,1}$&$\sigma_{B,1}$&$A_{A,0}$&$A_{B,0}$\\
    \hline
    2$\pi$ &  $2.27\pi$& 100& 100&2.5&2.84&1.4&1.4&0.75&0.75\\
    \hline
    \end{tabular}
    \caption{Parameters employed for the Fourier space two-point density-density correlation function in Eq. (\ref{eq:correlation})}
    \label{tab:Correlation_Params}
\end{table}

\section{{Numerical Implementation}\label{sec:Appendix_B}}

Dynamical evolution of Eqs. \ref{eq:n_dynamics} and \ref{eq:c_dynamics} in the main text was performed with a semi-implicit Fourier spectral method \cite{ProvatasBook}. 
The time discretized form we adopt for our equations in Fourier space is,
\begin{align}
    \frac{\hat{n}^{t+1}-2\hat{n}^t+\hat{n}^{t-1}}{\Delta t^2}+\beta\frac{ \hat{n}^{t+1}-\hat{n}^t}{\Delta t} &= \hat{L}_n^t \hat{n}^{t+1} -k^2 \hat{N}_n^t, \nonumber \\
    \frac{\hat{c}^{t+1}-\hat{c}^{t}}{\Delta t} &= \hat{L}_c^t \hat{c}^{t+1} -k^2 \hat{N}_c^t. 
\end{align}
Here, $\hat{L}_{n/c}^t$ and $\hat{N}_{n/c}^t$ are the linear and non-linear terms associated with each field in Fourier space and superscripts indicate the time-step of each term. In the above, we apply the frequently applied approximation that the linear operator associated with the evolution of the $n$ and $c$ fields is diagonal with elements $\hat{L}^t_{n/c}$. Rearranging the above, we can rewrite the discretized scheme as follows,
\begin{align}
\hat{n}^{t+1}&= \frac{\hat{n}^t(2 +\beta\Delta t ) -k^2 \hat{N}_n^t\Delta t^2 -\hat{n}^{t-1}}{(1+\beta\Delta t -\hat{L}_n^t\Delta t^2)} \\
\hat{c}^{t+1}&= \frac{\hat{c}^t -k^2 \hat{N}_c^t\Delta t }{(1 -\hat{L}_n^t\Delta t)}.
\end{align}
The fields can then be evolved using the information from the previous time-step for $c$ and previous two time-steps for $n$.
The parameters employed to evolve the dynamical equations are shown in Table \ref{tab:Dynamics_Params}. These parameters were employed to allow for simultaneous precipitation and plastic response for the strain-rates we employ. The choice of strain-rate was set such that a non-negligible amount of strain-energy could be imparted into the system prior to precipitation while staying within the confines of strains applied during deformation of materials. 

Each choice of temperature, strain-rate and activation energy were simulated using the same discretization scheme. We employ a two-dimensional system  box of $N_x\times N_y$ pixels with a discretization of $dx$ and $dy$ along each respective axis. Time was discretized with an increment $dt$. The parameters we employ for our simulations are shown in Table \ref{tab:Discretization_Params}. 

The initial discretizations ($dx^0$, $dy^0$) along each axis are varied along with the initial grain boundary misorientation to allow for an integer number of atoms initially present along both axes inside the simulation box. These variations are kept within the ranges of $12.00^{\circ}\pm 0.05^{\circ}$ and $0.10\pm 0.01$ for the misorientation and discretization respectively. We do not observe significant deviations in the behaviour precipitated fraction or the pressure over different parameter choices within our set tolerance (solutions are typically not unique within the tolerance range).

Application of deformation is performed numerically through periodic variation of the spatial discretization while preserving the overall volume of the system. The discretization is varied according to the following scheme,
\begin{align}
    dx^{t+T} &= dx^t(1+\dot{\epsilon}T dt) \\ 
    dy^{t+T} &= dy^t/(1+\dot{\epsilon}T dt)
\end{align}
where T denotes the number of numerical time-steps between each application of deformation.

\begin{table}[h]
    \centering
    \begin{tabular}{|c|c|c|c|c|c|c|}
     \hline
    $M_n^0$&$M_c^0$&$E_A^c$&$E_A^n$&$\beta$&$N_n$&$N_c$\\
    \hline
    0.02&1&0.1-1&10&0.1&0.05&0.01\\
    \hline
    \end{tabular}
    \caption{Parameters employed for dynamical evolution of Eqs. \ref{eq:n_dynamics} and \ref{eq:c_dynamics}.
    }
    \label{tab:Dynamics_Params}
\end{table}
\begin{table}[h]
    \centering
    \begin{tabular}{|c|c|c|c|c|c|}
     \hline
    $N_x$ & $N_y$ &  $dx$ & $dy$ & $dt$&$T$\\
    \hline
    4096&4096&$0.1\pm0.01$&$0.1\pm0.01$&0.5&300\\
    \hline
    \end{tabular}
    \caption{Numerical system size and discretization employed along the x and y axes. The initial grain boundary misorientation and discretizations vary slightly between simulations to accommodate an integer number of atoms parallele to the GB for varying supersaturation.
    }
    \label{tab:Discretization_Params}
\end{table}


\bibliographystyle{apsrev4-2}
\bibliography{Bibliography}

\end{document}